\documentclass[aps,pre,twocolumn,showpacs,groupedaddress,footinbib]{revtex4-2}

\usepackage{graphicx}
\usepackage{latexsym}
\usepackage{amsmath}
\usepackage{amsfonts}
\usepackage{amssymb}
\usepackage{color}

\usepackage{units}

\newcommand{\be}{\begin{equation}}
\newcommand{\ee}{\end{equation}}
\newcommand{\bea}{\begin{eqnarray}}
\newcommand{\eea}{\end{eqnarray}}

\usepackage[parfill]{parskip}

\begin{document}

\title{Reentrant Transitions in a Mixture of Small and Big Particles Interacting via Soft Repulsive Potential}

\author{Itay Azizi$^1$}
\email{aziziit@biu.ac.il}

\author{Alexander Y. Grosberg$^2$}
 \email{ayg1@nyu.edu}

\author{Yitzhak Rabin$^1$}
 \email{yitzhak.rabin@biu.ac.il}
\affiliation{$^1$Department of Physics and Institute of Nanotechnology and Advanced Materials, \\Bar-Ilan University,
Ramat-Gan 52900, Israel}
\affiliation{$^2$Department of Physics and Center for Soft Matter Research, New York University, NY 10003, USA}%

\begin{abstract}

We report the first observation of temperature-controlled reentrant transition in simulations of mixtures of small and big particles interacting via soft repulsive potential in 2D. As temperature increases, the system passes from a fluid mixture, to a crystal of big particles in a fluid of small particles and back to a fluid mixture. Solidification is driven by entropy gain of small particles which overcomes the free energy cost of confining big ones. Melting results from enhanced interpenetration of particles at high temperature which reduces the entropic forces that stabilize the crystal.

\end{abstract}


\maketitle

In most many-body systems ordering is produced by attractive forces overcoming the influence of entropy. An exception to this rule are systems with entropy-driven ordering characterized by an ordered phase with higher entropy than that of the disordered phase \cite{Onsager1949,Frenkel1994,Frenkel1999,Frenkel2015}. Thus, in lyotropic liquid crystals \cite{Onsager1949} nematic ordering arises as the result of interplay between rotational and translational degrees of freedom of elongated rigid molecules as the system gives up orientational entropy in order to gain translational entropy. A similar entropic mechanism was invoked in order to explain the observed surface freezing of alkanes \cite{Alexei1996}. Another example is the entropy-driven demixing transition in mixtures of small and big particles, that was originally observed in emulsions of silicon-oil-in-water stabilized by sodium dodecylsulfate (SDS) micelles \cite{Bibette1990}. Subsequently, this demixing phenomenon was demonstrated analytically in hard-core mixtures \cite{Hansen1991,Louis1992,Dijkstra1999,Tuinier2003}, in simulations \cite{Frenkel1996,Castenada2003,Cinacchi2007,Zaccarelli2015} and in experiments \cite{Pine1994,Leiderer1998,Pusey1995,Tuinier2003,Cook2006,Sapir2015,Briscoe2015}. The common understanding of this phenomenon is in terms of the entropic depletion mechanism proposed by Asakura and Oosawa \cite{AO1954,AO1958}: when the number of small particles is much larger than that of big ones, the  system  can  decrease  its  free  energy by creating a dense ordered phase of big particles, thus freeing space for small particles and therefore increasing their entropy. Note that this mechanism does not violate the second law of thermodynamics since the entropy decrease of big particles is overcompensated by the entropy increase of the small ones. For a recent review of this mechanism applications in a variety of fields see \cite{Zaccarelli2022}. 

Whereas this phenomenon is well-established, it raises an interesting issue that has not been addressed so far:  what happens to big particle crystals as temperature increases?  Clearly, for  hard  spheres there is no energy scale (the force between two particles changes abruptly from zero to infinity as they approach each other) and the phase diagram of the hard sphere mixture depends only on the size ratio and concentrations of the two components and not on temperature.  However, when the repulsive interaction between the particles is described by a continuous potential (such that the repulsive force is always finite), the characteristic distance of closest approach depends on temperature and the question about what happens at different temperatures remains open. The study  of  this  problem  is  the  subject  of  the  present  work which, for ease of visualization, is carried out in two dimensions.

\textit{Methods.} In order to elucidate the behavior of binary mixtures of big and small particles at different temperatures, we carried out Langevin dynamics simulations using LAMMPS (Large-scale Atomic/Molecular Massively Parallel Simulator \cite{LAMMPS}). Our two-dimensional system consists of $N_{\mathrm{s}}$ small and $N_{\mathrm{b}}$ big particles interacting via WCA (Weeks-Chandler-Andersen) potential given by $u_{ij}(r)=4\epsilon_{ij}[(\sigma_{ij}/r)^{12}-(\sigma_{ij}/r)^{6}]+\epsilon_{ij}$ with cut-off distance of $\sigma^{\prime}_{ij}=2^{1/6} \sigma_{ij} \approx 1.12\sigma_{ij}$, where $i$, $j$ are either $\mathrm{b}$ or $\mathrm{s}$. The LJ (Lennard-Jones) diameter of a big particle is taken to be ten times larger than that of a small particle: $\sigma_{\mathrm{bb}} = 10 \sigma_{\mathrm{ss}}$, and the sum rule is assumed for interactions between different particles, $\sigma_{ij}=(\sigma_{ii}+\sigma_{jj})/2$. Interaction energies are chosen to be all the same  $\epsilon=\epsilon_{\mathrm{ss}}=\epsilon_{\mathrm{bb}}=\epsilon_{\mathrm{bs}}$ (other values are used to prepare the initial mixtures - see below). The simulation box is a square of length $L=343\sigma_{\mathrm{ss}}$ with periodic boundary conditions in both directions. The LJ area fraction is defined by the cut-off diameter

\begin{eqnarray}
\phi_i=\frac{N_i \pi \left( \sigma^{\prime}_{ii}/2 \right)^2}{L^2}
\label{eq:LJareafraction} \ ,
\end{eqnarray}
with $i=\mathrm{b},\mathrm{s}$ (note that so defined ``area fraction'' can be greater than unity). The motion of particle $k$ is described by the Langevin equation
\begin{eqnarray}
m_k\ddot{\bf{r}}_k(t) = -\frac{\partial u_k}{\partial {\bf{r}}_k} -\zeta_k \dot{\bf{r}}_k(t) + \eta_k(t)
\label{eq:langevin} \ ,
\end{eqnarray}
where $m_k$ is the $k$-th particle mass, $u_k=\sum_j u_{kj}$ is the total potential energy associated with particle $k$ (the sum is over all its WCA interactions), $\zeta_k$ is the $k$-th particle friction coefficient and $\eta_k$ is a random thermal force with magnitude proportional to $(\zeta_k k_BT/\Delta t)^{1/2}$, where $k_B$, $T$ and $\Delta t$ are the Boltzmann constant, temperature and integration time step, respectively.

For simplicity we took the masses of big and small particles to be the same, $m_{\mathrm{b}}=m_{\mathrm{s}}$, and used it as a unit of mass (we verified that choosing masses to be different does not affect the statistical properties of our results, such as the average potential energy - not shown). We set the system of units by measuring all distances relative to $\sigma_{\mathrm{ss}}$, and all energies in units of  $\epsilon$; this automatically implies that all times are measured in units of the Lennard-Jones time defined as $\tau_{\mathrm{LJ}} = (m_{\mathrm{s}}\sigma_{\mathrm{ss}}^2/\epsilon)^{1/2}$. 

Following Stokes' law we set the friction coefficients $\zeta_k$ to be proportional to the diameters of the particles, $\zeta_{\mathrm{b}}=10\zeta_{\mathrm{s}}$. We chose the values of these friction coefficients such that the damping times, $\tau^{\mathrm{damp}}_k=m_k/\zeta_k$ (which determine the cross over from  ballistic  to  diffusive regime of motion due to collisions with molecules of the  implicit solvent) are $\tau^{\mathrm{damp}}_{\mathrm{b}}=5\tau_{\mathrm{LJ}}$ and $\tau^{\mathrm{damp}}_{\mathrm{s}}=50\tau_{\mathrm{LJ}}$.

Each simulation started from a well-mixed configuration obtained by setting $T=1$, low $\epsilon_{\mathrm{bs}}=0.1$, and high $\epsilon_{\mathrm{bb}}=\epsilon_{\mathrm{ss}}=2.0$; this choice of parameters promoted rapid mixing which took about $\sim20,000\tau_{\mathrm{LJ}}$. After a well-mixed state was reached, all the interaction parameters were replaced to $\epsilon_{ij}=1$ and the temperature was set to the required value $T$. At $T\leq1$ the time step was set to $\Delta t=0.005\tau_{\mathrm{LJ}}$ and at higher temperatures since the forces are stronger and we wanted to avoid numerical errors and/or displacements which are larger than the simulation box, the time step was decreased.

Since we are interested in equilibrium behavior, we waited until the systems relaxed into a steady state that was tested by three criteria: (1) relaxation to plateau values of each of the potential energy contributions $U_{\mathrm{bb}}(t)$, $U_{\mathrm{ss}}(t)$ and $U_{\mathrm{bs}}(t)$ ($U_{\mathrm{bb}}(t)$ is the total instantaneous potential energy contribution due to all bb interactions in the system, etc.), as a function of time, (2) observation of stationary nearest-neighbor distribution of big particles, and (3) movies of the relaxation of the system. The corresponding figures and movies are shown in Supplemental Material \cite{SupMat}.

\begin{figure}[ht]
\includegraphics[width=0.9\linewidth]{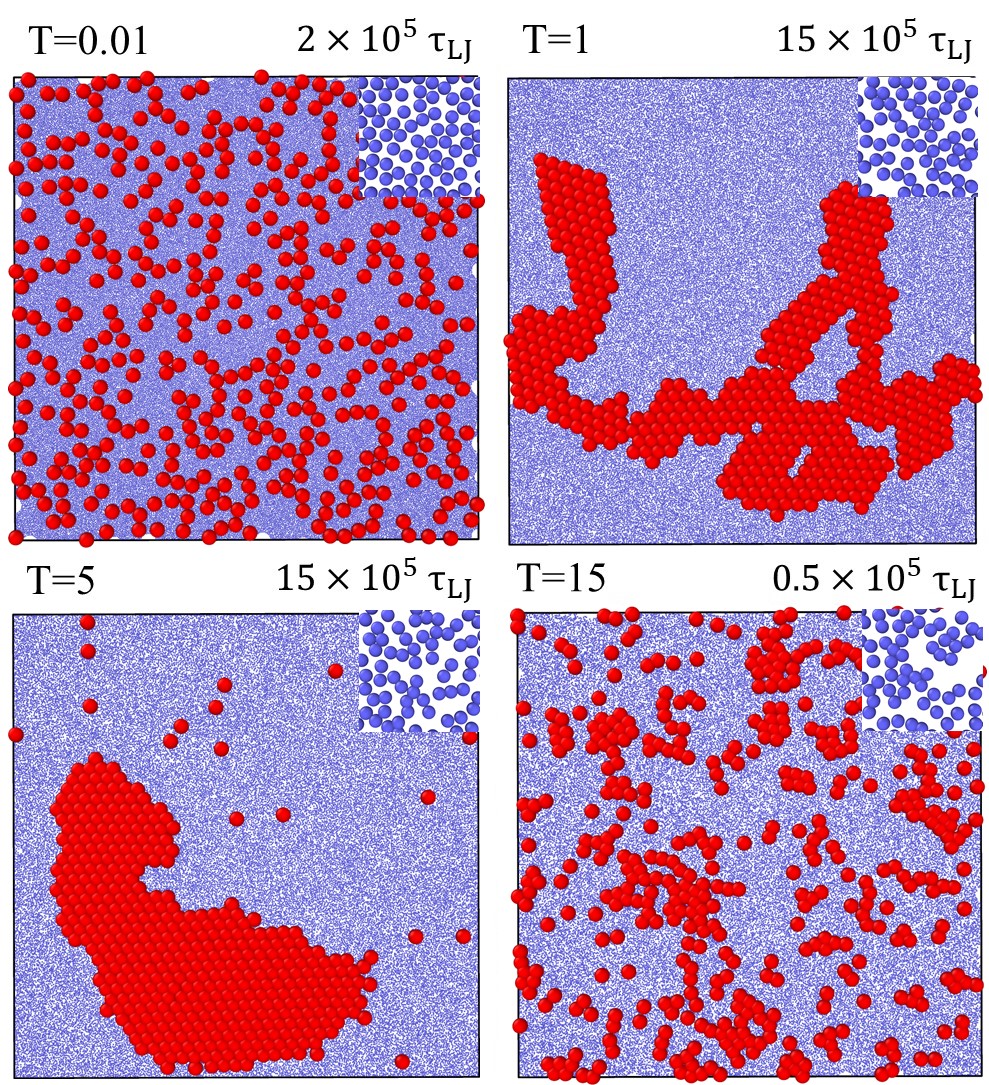}
	\caption{\label{fig:snapshots} Snapshots of a binary mixture with $N_{\mathrm{b}}=450$ ($\phi_{\mathrm{b}}=0.377$) and $N_{\mathrm{s}}=48690$ ($\phi_{\mathrm{s}}=0.408$) at $T=0.01,1,5$ and $15$. As temperature increases the system undergoes a reentrant transition from the uniform fluid mixture, to coexisting solid crystal of big particles surrounded by a fluid of small particles, and to another fluid mixture in which solid particles form small and not very long-living (labile) clusters. Note that the small particle fluid is zoomed in and presented in the corner of each snapshot.}
\end{figure}

\textit{Results: Description of snapshots and phase diagram.} In order to illustrate the reentrant behavior of the system upon change of temperature, in Fig. \ref{fig:snapshots} we present snapshots of a mixture of $N=49140$ particles, $N_{\mathrm{b}}=450$ ($\phi_{\mathrm{b}}=0.377$) and $N_{\mathrm{s}}=48690$ ($\phi_{\mathrm{s}}=0.408$), taken at several representative temperatures. At low temperature ($T=0.01$), a homogeneous fluid mixture of big and small particles is observed. As temperature increases the system undergoes demixing accompanied by formation of an elongated, branched and mostly hexatic solid of big particles (with few small particles inside, as defects), surrounded by a fluid of small particles ($T=1$). Upon further increase of temperature the solid crystal adopts a somewhat more regular shape (presumably because of increased surface tension) and begins to melt, releasing big particles into the small particle fluid ($T=5$). At yet higher temperatures a nearly homogeneous fluid mixture reappears, with many small unstable clusters of big particles surrounded by a fluid of small particles ($T=15$).

In what range of small and big particle concentrations does the reentrant behavior shown in the snapshots in Fig.\ref{fig:snapshots} take place? Notice that for the small particles to remain fluid at all temperatures, the total area fraction $\phi_{\mathrm{b}}+\phi_{\mathrm{s}}$ must not be too large (otherwise the kinetics would become frozen at low temperatures due to jamming).  

\begin{figure}[ht]
\includegraphics[width=0.9\linewidth]{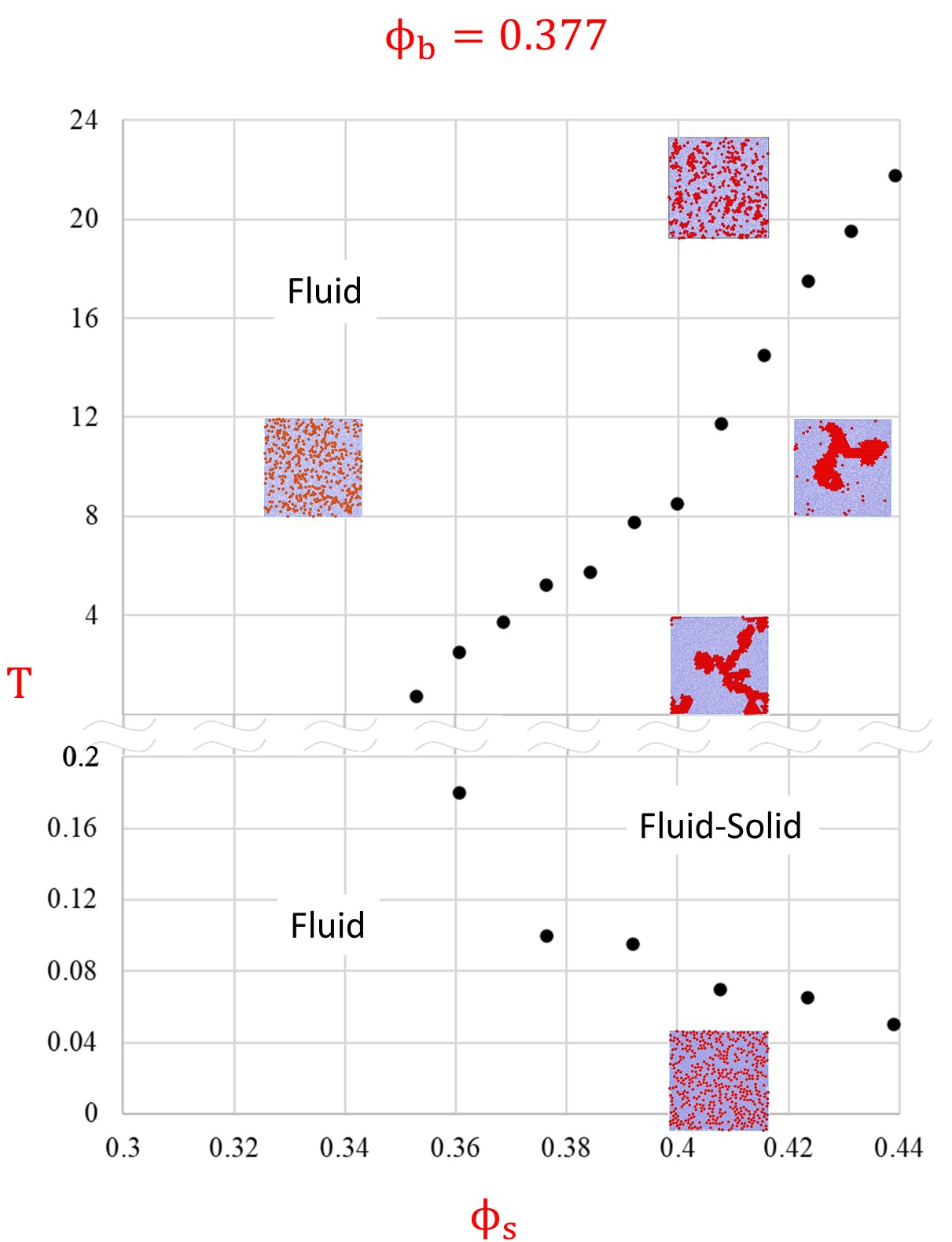}
	\caption{\label{fig:PD} Phase diagram of the system with $N_{\mathrm{b}}=450$ big particles  ($\phi_{\mathrm{b}}=0.377$) and temperature (in units of LJ energy $\epsilon$) as a function of area fraction of small particles $\phi_{\mathrm{s}}$ (defined based on the cut-off distance $\sigma_{\mathrm{s}}^{\prime}$, see Eq. (\ref{eq:LJareafraction})). Note the different temperature scales above and below $T \sim 1$.}
\end{figure}

For simplicity, we decided to use visual inspection of steady-state snapshots of the system (not shown) in order to construct an approximate phase diagram in the $\phi_{\mathrm{s}}-T$ plane, for a system with a fixed number of big particles $N_{\mathrm{b}}=450$ ($\phi_{\mathrm{b}}=0.377$)  (see Fig.\ref{fig:PD}).

The phase diagram shows that demixing accompanied by formation of a solid crystal of big particles (the Fluid-Solid region of the phase diagram) takes place only when the number of small particles exceeds a threshold value that corresponds to area fraction $\phi_{\mathrm{s}}^{\mathrm{crit}} \approx 0.35$. Notice that while below this value a homogeneous fluid mixture exists at all temperatures, a reentrant transition is observed for $\phi_{\mathrm{s}}>\phi_{\mathrm{s}}^{\mathrm{crit}}$.

Both coexisting phases in the Fluid-Solid region of the phase diagram, appear to be nearly pure in the sense that the fluid phase consists of mostly small particles (with only a few isolated large particles), and the solid crystal is composed of almost exclusively big particles (with a few small particles trapped inside voids). This motivated us to relate the two coexisting densities by running a separate auxiliary simulation on an one-component system and then equating the pressures of two approximately single-component phases.  As shown in the Supplemental Material \cite{SupMat}, this supports the assertion that the observed phases are well equilibrated.

What is the origin of the reentrant behavior observed in the phase diagram Fig.\ref{fig:PD}? We propose that the low temperature transition (lower branch of the phase diagram) from the Fluid to the Fluid-Solid region upon increasing the temperature, takes place via the entropic Asakura-Oosawa mechanism discussed in the introduction, once the entropy gain due to the increase of the area accessible to small particles due to condensation of big ones, overcomes the repulsive interactions between big particles.

However, this mechanism cannot explain the melting transition at yet higher temperatures (upper branch of the phase diagram), since the strength of the entropic forces that lead to the formation of solid crystal of big particles, is expected to increase with temperature and therefore stabilize the solid phase. In the following we will present a tentative explanation of this effect based on the analysis of radial pair distribution functions (RPDFs).

\begin{figure}[ht]
\includegraphics[width=0.9\linewidth]{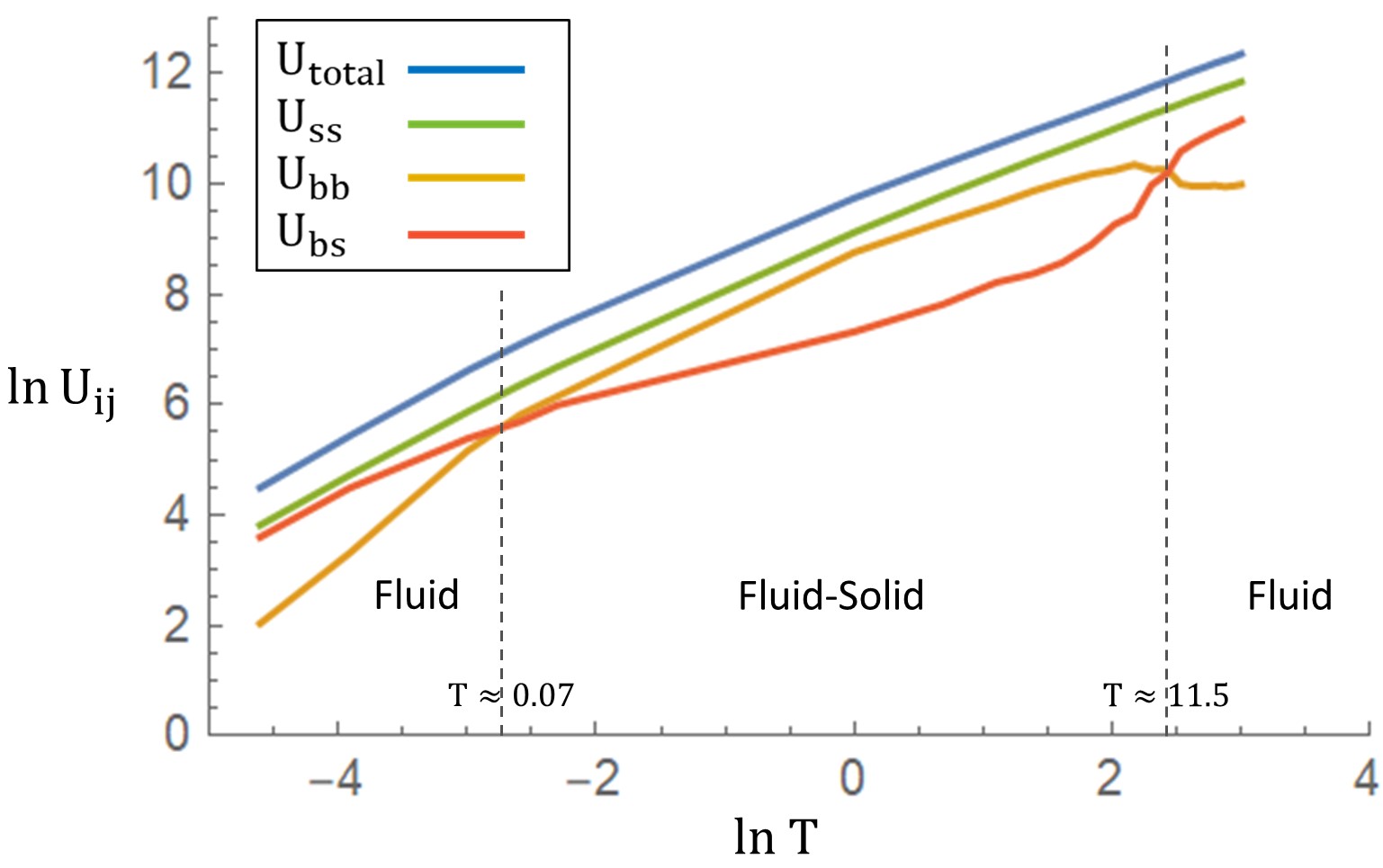}
	\caption{\label{fig:Ep} Potential energy contributions of the binary mixture of $N_{\mathrm{b}}=450$ big particles and $N_{\mathrm{s}}=48690$ small ones in steady state as a function of temperature (log-log scale).  The blue line shows the total potential energy $U_{\mathrm{total}}$. Green, orange, and red lines show potential energies of interactions between small particles $U_{\mathrm{ss}}$, between big particles $U_{\mathrm{bb}}$, and between big and small ones $U_{\mathrm{bs}}$, respectively. Coexistence of a solid of big particles with the fluid of small ones is characterized by $U_{\mathrm{bb}} > U_{\mathrm{bs}}$, this is observed, consistent with the phase diagram Fig. \ref{fig:PD} constructed by visual inspection of the system snapshots.}
\end{figure}

\textit{Results: Quantitative analysis.} In order to analyze the reentrant transition we focused again on the case of $N_{\mathrm{b}}=450$ big particles and $N_{\mathrm{s}}=48690$ small ones shown in Fig. \ref{fig:snapshots}, for which a large solid crystal of big particles was observed in the temperature range from about $T=0.07$ to slightly below $T=12$ (see Fig. \ref{fig:PD}). To supplement our visual observations of snapshots by a more quantitative method for determining the transition temperatures, we measured the different potential energy contributions as a function of temperature.  Inspection of Fig. \ref{fig:Ep} shows that $U_{\mathrm{bb}}>U_{\mathrm{bs}}$  in the temperature range $0.07-11.5$ (here, $U_{ij}$ is the total potential energy of interactions of all $ij$ pairs in the system). Since the appearance of a solid crystal must be accompanied by a large increase in the number of $\mathrm{bb}$ interactions, this supports our visual determination of the boundaries of the Fluid-Solid region in the phase diagram.

To gain further insight into pair contacts between particles, and to particles ability to climb one-another's potential barriers, we proceed to analyze the temperature dependence of the  radial pair distribution functions (RPDFs) $g_{\mathrm{ss}}(r)$, $g_{\mathrm{bb}}(r)$ and $g_{\mathrm{bs}}(r)$. While higher order correlation functions are needed to completely characterize liquids and solids, RPDFs contain important information about the state of the system. For example, visual inspection of a diagonal ($\mathrm{bb}$ or $\mathrm{ss}$) RPDFs can distinguish between a liquid (1-2 peaks) and a solid (multiple peaks), and the position and the amplitude of the first peak inform us about the positions and the number of particles in the first coordination shell surrounding the chosen particle.

\begin{figure}[ht]
\includegraphics[width=0.7\linewidth]{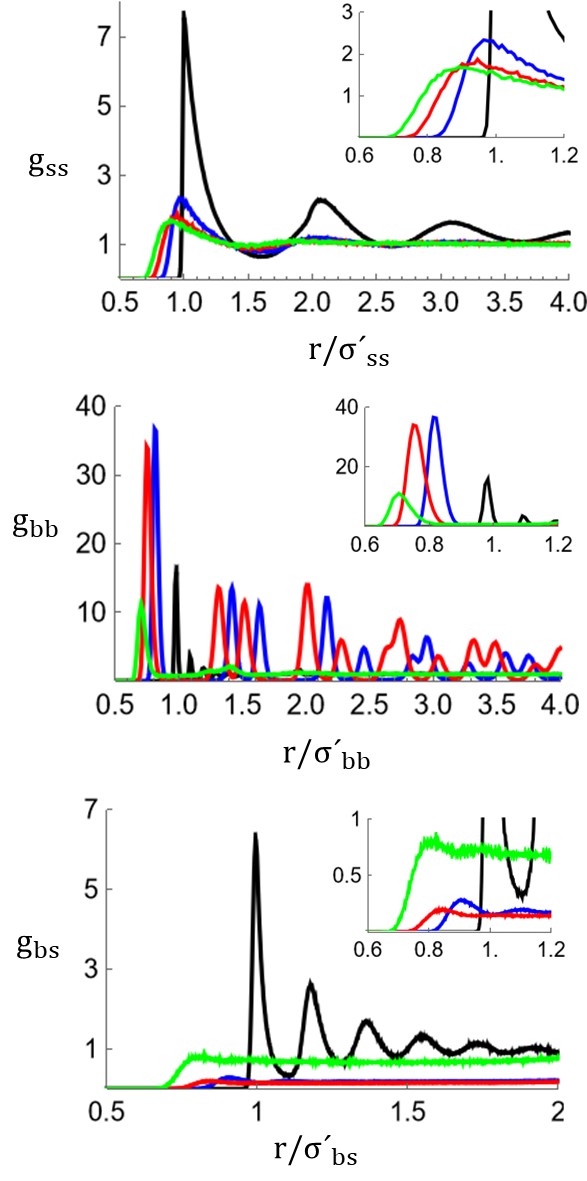}
	\caption{\label{fig:RDFs} Small-small, big-big and big-small radial pair distribution functions (RPDFs) of the binary mixture with $N_{\mathrm{b}}=450$ big particles and $N_{\mathrm{s}}=48690$ small ones at $T=0.01$ (black), 1 (blue), 5 (red) and 15 (green), are plotted as a function of distance. For each RPDF, $g_{ij}$, the distance is scaled with the respective cutoff distance, $\sigma_{ij}^{\prime}=2^{1/6} \sigma_{ij}$. In the insets, we zoom in the functions and show that for all types of RPDF, the position of the first peak decreases with increasing temperature.}
\end{figure}

Indeed, inspection of Fig. \ref{fig:RDFs} shows that the RPDF of the small particles $g_{\mathrm{ss}}(r)$ is characteristic of a fluid: a well-defined first peak and a weak second peak are observed at all temperatures. As expected from our snapshots in Fig.\ref{fig:snapshots}, the distribution function of big particles is  fluid-like (a pronounced first peak and strongly suppressed subsequent peaks) at both high and low temperatures, while at intermediate temperatures ($T=1$ and $T=5$)  $g_{\mathrm{bb}}$ shows the characteristic features of a hexatic solid: a well-defined primary peak and a split second peak, followed by additional peaks in the range shown in Fig.\ref{fig:RDFs}). Examination of the amplitude of the first peak of $g_{\mathrm{bs}}$ indicates that the number of $\mathrm{bs}$ contacts is maximal at $T=0.01$, very low at $T=1$ and $T=5$ and intermediate at $T=15$, again in agreement with expectations based on the snapshots in Fig.\ref{fig:snapshots}.

While the above observations provide further support to the  qualitative results reported in the previous subsection, the realization that the position of the first peaks of all RPDFs decreases as a function of temperature, provides new and important information. The fact that the first peak moves to smaller distances as temperature increases indicates that the typical separation between neighboring particles decreases with temperature increase. This concurs with the intuitive expectation that the probability that particles can penetrate into regions of high repulsive potential and approach each other, increases with temperature.

Based on the above observations we propose a mechanism of high temperature melting of solid crystals in our system. Notice that while the distance of closest approach of hard core particles does not change with temperature,  the probability that particles which interact via a soft repulsive potential (such as WCA) can come increasingly close to each other, grows with temperature. This effect increases the volume available to the particles and therefore decreases the entropic depletion forces that led to demixing and to the formation of solid crystals of big particles. We conclude that the suppression of the Asakura-Oosawa mechanism at elevated temperatures is responsible for the observed melting transition.

\textit{Discussion.} In this work we used computer simulations in two dimensions to study the effects of temperature on formation and melting of solid crystals in highly asymmetric binary mixtures of small and big particles interacting via a soft repulsive WCA potential. Using visual inspection of snapshots of the system taken at different temperatures we found that the system undergoes a reentrant transition with increasing temperature, from a homogeneous fluid mixture at low temperatures, to a crystal solid of big particles surrounded by a fluid of small particles at intermediate temperature. Upon further increase of temperature, the crystal melts resulting in formation of small aggregates of big particles mixed in a fluid of small particles. Based of the analysis of the temperature dependence of the RPDFs, we argued that the high temperature melting transition arises from growing interpenetration of particles at elevated temperatures that increases the available volume and suppresses the entropic depletion forces that stabilize the solid phase.

Reentrant mixing-demixing transitions are well-known in polymer solutions and blends that have both lower and upper critical solution temperatures \cite{Flory1953}, and more recently they have been observed in colloid-polymer mixtures \cite{Chaikin2015}. In computational models, similar reentrant behavior was noticed for a system with artificial so-called Jagla (piece-wise linear) potentials \cite{Buldyrev2007}.  

The present paper reports the first observation of such phenomena in a simple model system: a mixture of small and big particles that interact via purely repulsive forces. The simplicity of our system allowed us to propose a simple and intuitive mechanism for both the formation of crystals and for their melting, with increasing temperature.

Our interpretation of the reentrant crystal melting, based on imagining effective reduction of the excluded volume by particles climbing high up the smooth repulsive potential, implies that phase behavior of the system should be rather sensitive to the specific profile of the interaction potential $u(r)$. It would be an interesting challenge to formulate the quantitative conditions on $u(r)$ necessary to realize the reentrant behavior. Another twist on the same question is how to engineer the $u(r)$ to realize the desirable phase behavior. This question is potentially a very practical one, given the diversity of possible interactions in colloidal systems, including electrical double layers, mediation by dissolved and grafted polymers, and many others \cite{Weitz2017, Kreiner:2021}.

We would like to mention some of the limitations of the present study. First, our simulations were done in two dimensions and all the reported results refer to a system of the size $L = 343 \sigma_{\mathrm{ss}}$. This system size was chosen to allow us to scan a broad range of densities and to achieve steady states in a reasonable computation time. Given the transparent physical picture that emerged from our study, we fully expect that qualitatively similar behaviour for larger systems, and both in two and three dimensions. Second, for our simulations we examined small and big particles with size ratio ($\sigma_{\mathrm{ss}} / \sigma_{\mathrm{bb}}$) of 0.1. In general, the size ratio is an important parameter controlling the phase behavior of a binary mixture. Since phase separation is obviously absent when size ratio is equal to unity, one may ask what the maximal size ratio is to observe phase separation, and then to observe the reentrance. This maximum corresponds to a critical point.  Finding this point and exploring its vicinity is an interesting challenge, although perhaps a steep one, given the expected slowdown of relaxation around critical point. This problem is squarely beyond the framework of the present paper, we are interested in the regime far beyond critical.  In this sense, specific choice of the size ratio, as long as it is far from the critical, does not really matter as far as qualitative features are concerned, although of course specific range of concentrations for phase separation does depend on the size ratio.

\section*{Acknowledgments}
IA acknowledges many valuable discussions with Giorgio Cinacchi and the use of HPC facilities of NYU. AYG acknowledges stimulating and enjoyable discussions with Sergey Buldyrev. YR would like to acknowledge support by grants from the Israel Science Foundation (178/16) and from the Israeli Centers for Research Excellence program of the Planning and Budgeting Committee (1902/12). \newline

\bibliography{ref}

\end{document}